**Instability-driven mechanically locked states in functional oxide membranes**

Varun Harbola[1*†], Thomas Emil le Cozannet[2*], Denis Alikin[3], Shinhee Yun[2], Edwin Dollekamp[2], Andrea Roberto Insinga[2], Rasmus Bjørk[2], Nikolas Vitaliti[2], Thomas Sand Jespersen[2], Katja Isabelle Wurster[2], Dae-Sung Park[2], Jochen Mannhart[1], & Nini Pryds[2‡]

[1] Max Planck Institute for Solid State Research, Heisenbergstrasse 1, 70569 Stuttgart, Germany

[2] Department of Energy Storage and Conversion, Technical University of Denmark, Kongens Lyngby 2800, Denmark

[3] Department of Physics, University of Aveiro, Campus Universitario de Santiago, 3810-193 Aveiro, Portugal

*These authors contributed equally to this work.

Author to who correspondence should be addressed: †v.harbola@fkf.mpg.de and ‡nipr@dtu.dk

**Abstract:**

Mechanical instabilities in thin solids offer a powerful route to engineer nonlinear responses, yet their controlled use in functional crystalline oxides has remained largely unexplored. Notably, by changing the aspect ratio of solids, the energy landscape around equilibrium can be modified to induce non-linearities under lateral stresses through non-lateral deformations. These nonlinear systems can develop multiple local energy minima where the system can settle and switch between states through the application of a driving force. Crucially, recent advances in oxide thin film growth have enabled the fabrication of freestanding oxide membranes, paving a viable path for their use in bistable architecture, particularly at the nanoscale. Here, we demonstrate that freestanding oxide membranes, such as $SrTiO_3$ (STO) and $BaTiO_3$ (BTO), relax into well-defined metastable buckling states when transferred onto lithographically defined cavities. The membrane deformation is determined by the interplay between built-in residual strain, bending stiffness, and cavity geometry, resulting in reproducible bistable states with distinct strain distributions. Using a combination of atomic force microscopy, in-contact Kelvin probe measurements, and finite-element modelling, we reveal that these mechanically locked states directly shape the electromechanical potential landscape of ferroelectric $BaTiO_3$. We further demonstrate reversible snap-through transitions between mechanically degenerate states, establishing complex oxides as deterministic, geometry-tunable building blocks for nonlinear nanoelectromechanical architectures. Our results illustrate a general strategy for exploiting mechanical instabilities to encode and manipulate functional responses in ultrathin crystalline membranes.

**Main:**

**Introduction:** Mechanical instabilities such as buckling, snap-through, and multistability are central to a wide range of natural and engineered systems, enabling large mechanical responses to small stimuli and providing access to nonlinear energy landscapes (*1*, *2*).When deliberately engineered, these instabilities furnish powerful mechanisms for actuation, sensing, energy harvesting, and programmable mechanics. A classic example is the out-of-plane Euler instability in geometrically confined mechanical systems(*3*), in out-of-plane buckling occurs under in-plane compressive stress (*4*). These nonlinear systems are characterized by multiple local energy minima, which allow the system to adopt different metastable states. Switching between such states is achieved through applying an external driving force (*5*, *6*). The simplest of these configurations exhibits two discrete stable states, each corresponding to a unique geometric configuration, and requires a snap-through force to actuate the transition between them (Fig. 1a) (*5–7*). High-efficiency smart actuators (*8*), energy harvesters with optimal operational characteristics (*9*, *10*), flexible soft actuators for robotics (*11*, *12*), high-efficiency energy absorbers(*13*), and geometrically programmable materials (*14*) collectively demonstrate that mechanical bistability, when paired with thoughtful geometric and material design, has significantly advanced multiple fields. Extending such concepts to functional crystalline oxides, materials, in which electronic, structural, and ferroic properties are tightly coupled to lattice distortions (*15*), could unlock new device modes in which mechanical instabilities act directly on ferroelectric, flexoelectric, or magnetic degrees of freedom (*16*, *17*). This provides operational regimes where the functional response shows snap-through variations upon switching of the bistable states (*18*). Therefore, oxides represent a highly fertile class of materials to couple bistable mechanical modes with functional properties. However, the intrinsic brittleness and high stiffness of oxide crystals have historically prevented achieving large deformations required to access nonlinear mechanical regimes.

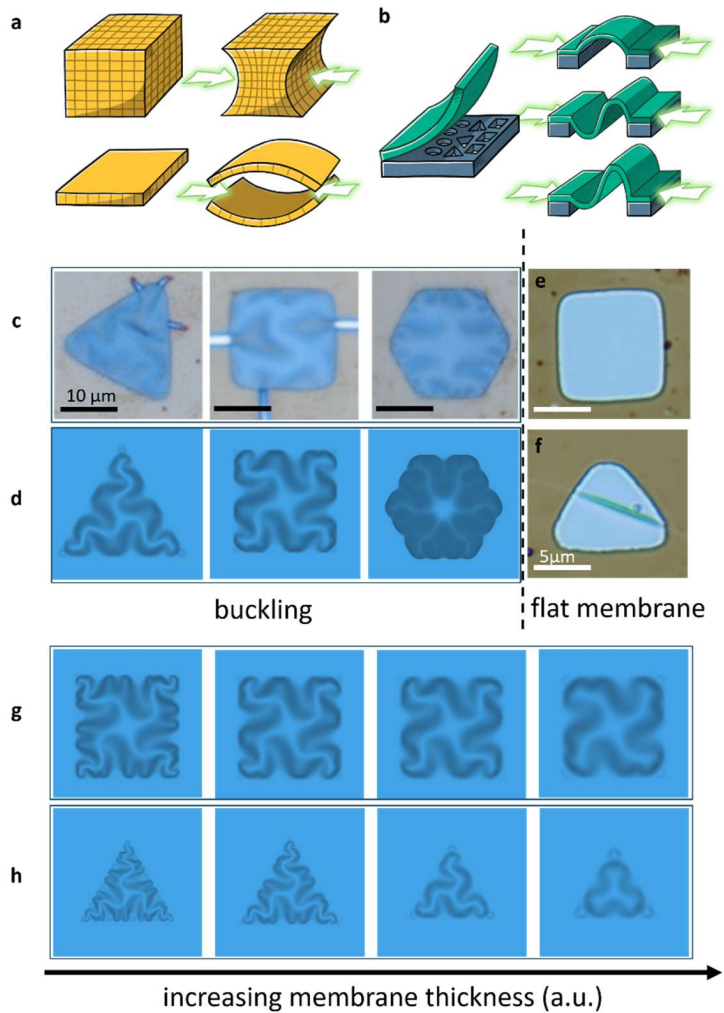

**Fig. 1: Deformation and buckling of suspended membranes. a,** A 3D structure, when subjected to external forces, deforms in a manner governed linearly by the magnitude of the force and the elastic properties of the material. When the aspect ratio of the structure is decreased, a buckling instability can be induced, which results in an out-of-plane buckling upon the influence of an in-plane force, a process that is highly non-linear and introduces a snap-through deformation. **b,** The buckling instability can also be introduced by transferring a thin membrane of a material which is in a zero-strain condition and allowed to relax to the zero stress condition over cavities. Depending on the size and shape of the cavity, the buckling can even have a curvature inversion and break the degeneracy between an up and down buckled membrane. **c,** Optical images of 10 nm thick $SrTiO_3$ membranes transferred onto cavities of different shapes. All membranes appear to be randomly wrinkled but are actually showing predictably states buckled with a higher order deformation. **d**, Simulated buckled membranes on variously shaped cavities. The buckling behavior of the membrane can be modelled well by allowing a membrane with zero strain to relax and reach a zero-stress state with finite element modelling. The simulated buckled profiles match well with the experimentally observed shapes. **e, f,** When the size of the hole is reduced, the threshold stress required to go through the buckling instability increases, resulting in membranes with a propensity to remain flat on the cavities. **g, h,** Simulation of thickness dependent buckling of membranes. To reliably control curvature and buckled states, it is essential to understand how membrane thickness influences buckling. Simulations qualitatively show that while keeping the size of the cavity the same (with all units self-consistent within the finite element model), by increasing the thickness of the membrane, the overall buckling and the amount of 'wrinkling' in the membrane can be controlled and reduced. The trend is independent of the shape of the cavity and can be understood in a straightforward manner by understanding that the bending stiffness of a membrane scales as the third power of its thickness, thus increasing the energy cost of buckling drastically with increasing thickness.

Complex oxides exhibit a variety of physical phases—ranging from magnetic and polar to superconducting—which are usually strongly coupled to the underlying lattice(*15*). Consequently, oxides represent a fertile class of materials to couple bistable mechanical states with functional properties. However, their inherent brittleness makes it challenging to sustain high strain, while their high bulk Young's modulus typically demands impractically large forces to achieve mechanical switching.

Recent advances in the synthesis of freestanding single-crystal oxide membranes(*19–22*) now alleviate these constraints by suppressing substrate clamping and drastically reducing the effective bending stiffness at nanometre thicknesses (*23*, *24*). Such membranes can sustain unusual flexibility(*25–28*) with a thickness-dependent Young´ modulus (*29–32*) while preserving their ferroic and electronic functionality for nanotechnological applications (*33*). These developments raise an important question: Can ultrathin oxide membranes be driven into well-defined mechanical instabilities, and if so, can these instabilities be exploited to engineer functional electromechanical states?

Here, we answer this question by demonstrating deterministic buckling and bistability in freestanding $SrTiO_3$ and $BaTiO_3$ membranes suspended over lithographically defined cavities. We show that the built-in strain of as-grown films, once released from their sacrificial layers, drives out-of-plane buckling whose symmetry and metastability are dictated by cavity geometry and membrane thickness. Using atomic force microscopy and finite-element analysis, we identify distinct stable and metastable deformation modes, each associated with characteristic strain fields. In $BaTiO_3$, these mechanical states give rise to strongly spatially modulated electromechanical potentials, enabling direct coupling between mechanical instabilities and ferroic functionality. Finally, force–displacement spectroscopy reveals reversible snap-through transitions between the mechanically locked states, establishing complex oxides as a versatile, geometry-programmable platform for nonlinear nanoelectromechanical systems.

**Transfer and buckling of complex oxide membranes:** Freestanding STO and BTO membranes were fabricated by epitaxial growth of the oxide on a water-soluble $Sr_3Al_2O_6$ (SAO) sacrificial layer,(*34*) followed by polymer-supported lift-off and transfer onto silicon chips patterned with ~10 μm cavities etched through deep reactive ion etching steps (Fig. 1b) (Methods). Upon polymer removal, the membranes become mechanically isolated and relax the in-plane strain accumulated during growth due to the lattice mismatch between the SAO sacrificial layer and the STO and BTO. This relaxation induces out-of-plane buckling whose morphology depends strongly on cavity geometry and membrane thickness (Fig. 1c–h). Larger cavities allow the residual strain to exceed the critical buckling threshold, producing higher order buckling modes. In contrast, for small cavities (≤10 μm), membranes often remain flat (Fig. 1e, f), indicating that their critical buckling load is not surpassed. Finite-element modelling of square and triangular holes (See the Methods) (Fig. 1d) captures these transitions and explains their scaling through classical Euler instability arguments: the critical load scales as $t^3/L^2$, where *t* is the membrane thickness and $L$ is the characteristic cavity size of the freestanding membrane (*35*, *36*). This scaling shows that reducing cavity size below a certain threshold prevents growth-induced stress from exceeding the critical level required for out-of-plane buckling in membranes of the same thickness. Consequently, the membrane maintains a zero-strain state instead of transitioning to a buckled, zero-stress configuration. This phenomenon is confirmed experimentally as the cavity size (*L*) decreases. Additionally, the scaling indicates that increasing *L* results in greater instability, leading to structures that appear randomly wrinkled. However, these structures represent higher-order structured buckling modes as a consequence of strain relaxation, and their geometry can be accurately predicted numerically based on the relevant residual stresses and boundary conditions. The nature of buckling, the number of undulations in the buckled states, and the lateral size over which the membrane buckles are all directly influenced by the membrane thickness when the cavity size and shape are held constant (Fig. 1g, h).

**Strain generation, potential landscape and metastability in ferroelectric membranes:**

The large curvature of BTO membranes generates local electromechanical potentials arising from flexoelectric polarization, strain-modulated ferroelectric polarization, and electrostatic interactions in the AFM tip–sample junction. To probe this, we performed piezoresponse force microscopy (PFM) and in-contact Kelvin probe measurements (KPFM) on both clamped and suspended membrane regions (Supplementary Information Figs. S1 and S2). For BTO membranes, we identified an optimal parameter space—a "Goldilocks zone"—at a specific thickness (30 nm) and the lateral size of the cavity (10 μm). In this optimized regime, the out-of-plane buckling exhibits low-order deformations with amplitudes reaching hundreds of nanometers. Crucially, the resulting buckling is more uniformly distributed across the freestanding area, contrasting with the largely localized deformations typical of higher-order, wrinkle-like modes (Supplementary Information Figs. S3-S5). In general, the electromechanical signal in PFM represents a combination of the true piezoresponse of the sample and the electrostatic force acting on the probe (*37*). In the regions of the substrate without cavities, the signal is predominantly piezoelectric in nature. However, on suspended buckled membranes, the ratio of piezoelectric to electrostatic contributions depends strongly on the mechanical boundary conditions, particularly the stiffness and constraints of the membrane (Supplementary Information Figs. S3-S5). To clarify the origin of the PFM signal in each case, the position of resonance bands in the cantilever's frequency spectrum was analyzed (see Supplementary Information Fig. S3-4). This analysis confirmed that the signal measured on the membrane is dominated entirely by the electrostatic force. In-contact electrostatic measurements can be used to extract information about the distribution of built-in potential in ferroelectric materials(*37*). Since the signal on the membrane is dominated by electrostatic interaction, the piezoelectric contribution can be neglected, and the in-contact surface potential can be determined by applying superimposed AC and DC voltages and identifying the DC bias that minimizes the electromechanical response, analogous to open-loop KPFM measurements (*38*).

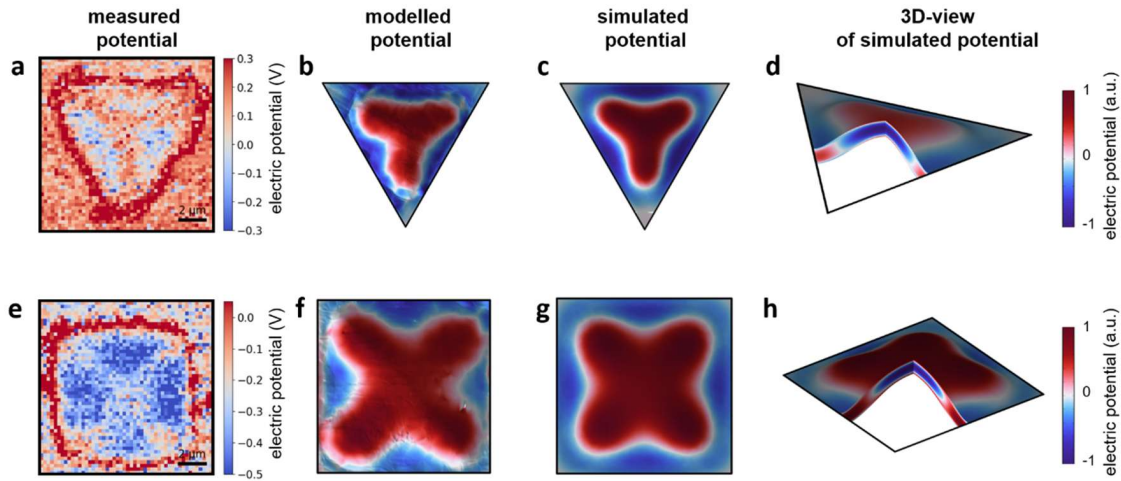

**Fig. 2: In-contact surface potential measurements and simulations. a, e,** Measured in-contact surface potential, **b, f,** Modelled potential from experimental displacement. **c, g,** Potential from the simulated finite element displacement and **d, h,** cross-section of the potential distribution of which the surface potential is show in (c) and (g). The large enhancement of the surface potential on the boundary of the membranes in (a) and (e) is an artefact, appearing as a result of the imperfect tip contact with the membrane in these regions. The qualitative consistency between simulated and measured surface potential, along with the buckled deformation, is remarkable and denotes a high degree of predictability for engineering electromechanical states in freestanding membranes. Scale bars represent all calculated potentials.

The results of the built-in surface potential measurements are presented in Fig. 2 (and further in Supplementary Information Fig. S6), where they are compared with numerical simulations made by the finite element method. The simulation results both from displacement obtained via the suspended topography and through simulated displacements show a clear correspondence to the experimental potential distribution, suggesting that mechanical strain plays a dominant role in shaping the observed surface potential. Additionally, simulations allowed us to visualize the volumetric distribution of the surface potential (Fig. 2d, h). It is to be noted that the simulation results presented are all in the self-consistent finite element units with only piezoelectric contributions, and are hence presented in arbitrary units. Furthermore, the experimental potential of the square hole is taken for a down buckled state but the simulation result is shown for an up buckled state for visual cavity as in the simulations, the degeneracy between up and down buckled states is still maintained. For a true quantitatively accurate model, the full elastic, ferroelectric and flexoelectric tensors need to be used, which also need to be characterized for $BaTiO_3$ in this thickness regime, an endeavor that is beyond the scope of this work. These results confirm that mechanically imposed curvature produces predictable electromechanical states, providing a design pathway for mechanically reconfigurable polarization landscapes.

Mechanical bistability emerges naturally from the degeneracy of up- and down-buckled configurations in thin membranes. Surprisingly, we find that cavity geometry seems to breaks this degeneracy. Square cavities preferentially stabilize downward buckling which made it difficult to search for an upward buckled membrane for the flipping experiment, whereas triangular cavities seem to favor upward buckling and upon force cycling return to an up buckled state (Fig. 3). This symmetry breaking likely results from asymmetries at the cavity edge or from slight differences in residual strain distribution. However, this bias does not qualitatively affect the metastability of both states, and upon an application of force, the state of the membrane can be switched between the two distinct buckled states. The force response clearly shows non-linearity, a snap-through instability, and characteristic hysteresis resulting from spontaneous switching between the two bistable states after crossing a certain force threshold (Figure 3). The force required to flip the membrane is similar (≈700 nN) and independent of the hole shape, both for the square and the triangular. However, what is intriguing is the shape and magnitude of the hysteresis, which differ significantly between the two dissimilarly shaped holes. This reflects a clear preference between the two buckled states of the membrane. For the triangular holes, the hysteresis is significantly narrower, and the membrane snaps back to the upward-buckled state even when the tip is still forcing it down, indicating a distinct bias towards the up-buckled state. In contrast, for a membrane on a square hole, the hysteresis is wider, and only a single snap-through event is observed throughout the force-displacement loop. After applying the force, the membrane has flipped from an up-buckled state to the down-buckled state, and the adhesive force between the tip and the membrane is insufficient to revert it back to the up-buckled state. This behavior is repeatable, and the membrane does not yield or break upon cycling through repeated application of the force (Supplementary Information Fig. S7). These two distinct characteristics, depending on the hole shape, illustrate how the metastability can be tailored to particular applications of the bistable/metastable architecture.

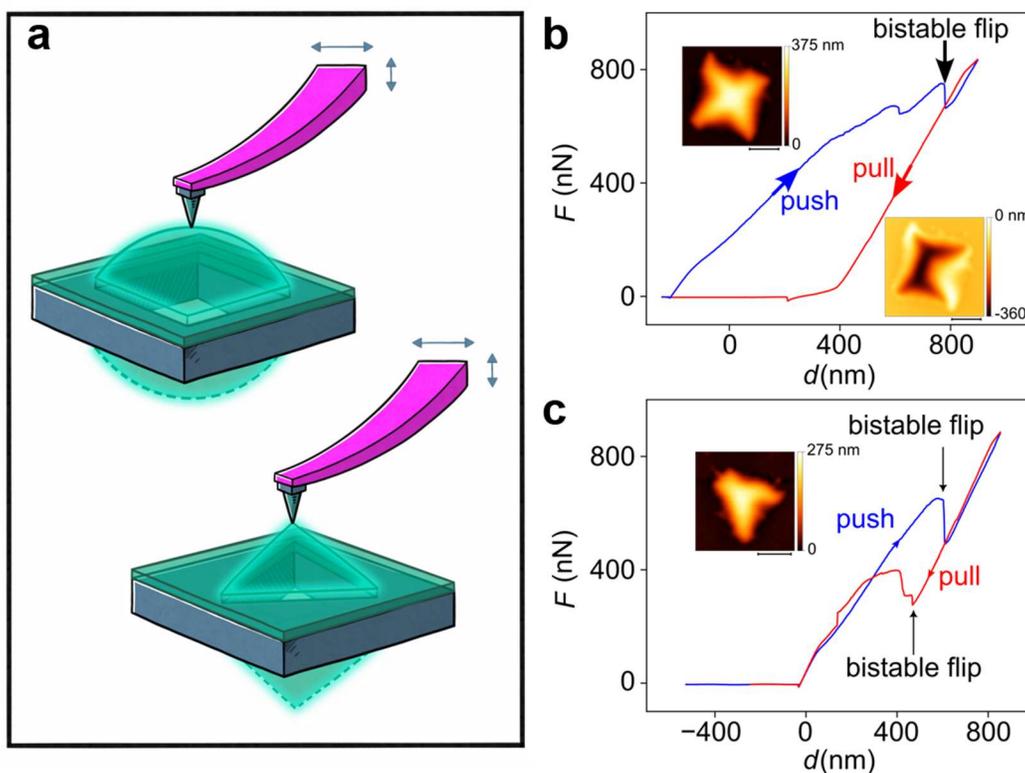

**Fig. 3: Bistability of buckled membranes. a,** Schematic representation of measuring the mechanical response of membranes suspended across cavities of different shapes. An AFM tip is used to press on a buckled membrane, and the deflection of the tip indicates the force being exerted by the membrane. By moving the tip vertically by a distance $d$, the $F$-$d$ response of the membrane can be recorded to study the metastability of the buckled state. **b,** $F$-$d$ response of a membrane on a square cavity. The cavity is buckled up to initially. As the membrane is pushed, the membrane flips into a down-buckled state. Before and after states of the membrane are displayed as insets (scale bar is 5 μm) and show the topography of the complementary up and down buckled state. **c,** $F$-$d$ response of a membrane on a square cavity. The cavity is buckled up to start with. As the pushing force is increased, the membrane undergoes a transition to a down-buckled state; however, upon retraction, it flips back to an up-buckled

state. The snap through events in both cases of the square and circular holes are sharp and showcase a distinct threshold non-linearity. The hysteresis curve in the case of a triangular hole is much narrower and the deformations show a bias to an up buckled state. Note the small sharp elbows in the *F-d* response. They denote the other smaller metastable states the system passes through before a macroscopic flip of the buckled state. The buckles state is shown as an inset where the scale bar denotes 5 μm.

Having two accessible stable states, each associated with its own strain distribution determined by deformation, provides a flexible solution for engineering strain-driven responses in these membranes. Due to the strong electromechanical coupling in BaTiO$_3$, strain, for example, shifts the piezo response coefficient, up to first order via a Taylor expansion around the relaxed state as:

$$d_{ijk}(\epsilon) = d_{ijk}(0) + \frac{\partial d_{ijk}}{\partial \epsilon_{mn}}\epsilon_{mn} + O(\epsilon^2) \tag{1}$$

Where $d_{ijk}$ is the piezoelectric tensor and $\epsilon$ is the induced strain. It furthermore may even stimulate local reorientation of the spontaneous polarization(*39*). This modification opens up new opportunities for versatile handling of electromechanical systems utilizing functional materials, which can be manipulated between two or more distinct states.

Our results establish freestanding single-crystal oxide membranes as a platform for accessing and controlling nonlinear mechanical states in functional materials. When released from their growth substrate and suspended over designed cavities, STO and BTO membranes relax their built-in residual strain through large out-of-plane buckling, resulting in stable or metastable shapes that can be engineered through geometry and thickness. These mechanically locked configurations are highly reproducible and serve as deterministic templates for strain-driven functional responses.

In ferroelectric BTO, the buckled states generate strongly spatially varying electromechanical potentials that arise from strain gradients and polarization–strain coupling. The strong agreement between measurement and finite-element predictions demonstrates the predictability of these mechanically encoded electromechanical states. Because many ferroic and electronic properties in oxides are strain-sensitive, from ferroelectric polarization to correlated electronic phases, the ability to imprint strain landscapes through mechanical instabilities opens up routes to new device modalities.

The demonstration of reversible snap-through transitions further suggests potential functionalities such as mechanically reconfigurable ferroelectric elements, bistable nanoactuators, strain-programmable memory, and nonlinear NEMS resonators. Integrating electrodes, domain-engineered ferroelectrics, or multilayer oxide heterostructures with such buckling architectures could provide dynamic control over ferroic states and enable multifunctional electromechanical devices.

More broadly, this work provides a foundational blueprint for coupling mechanical instabilities with functional properties in crystalline oxides. By leveraging geometry and thin-film elasticity, mechanical bistability becomes a design tool for manipulating nanoscale strain and its coupled electronic or ferroic responses.

**Methods:**

**Substrates:** The fabrication of the holed substrates involves 6 main steps on 100mm double-sided polished <100> n-doped Si wafers (350 μm): (1) The wafer's backside is patterned with 4μm AZ MiR 701 resist via UV lithography, creating 3x3mm windows and alignment marks using remover 1165 as developer. (2) A high etch rate DREM process is used with a 13.56 MHz supply. The deposit step flows C4F8 at 200 sccm (25 mtorr, 2 kW coil, no platen, cycle 4s). The removal step flows $SF_6$ and $O_2$ (350 sccm/35 sccm, 20 mtorr, 2.8 kW coil, 130 W platen, 1.5 s). The etch step uses similar parameters but at 100 mtorr pressure and 40 W platen for 5.5 s, repeated for 102 cycles to thin areas to >50 μm. The resist is then removed with oxygen/nitrogen plasma at 1 kW. (3) The wafer is coated with 50 nm Al2O3 via ALD, using TMA (150 sccm, 0.1s pulse) and $H_2O$ (200 sccm, 0.1s pulse) at 200°C. (4) A second photoresist mask (AZ 5214E) with 10 μm holes is spun and patterned, aligned over the thin Si areas with remover 1165. (5) The backside is protected with polymer tape, then the wafer is dipped in BHF (12% HF with ammonium fluoride) for 70 s at 20°C. The tape is removed, and the resist is stripped with oxygen plasma. (6) The holes are etched using DREM, with $Al_2O_3$ as an etch stop.

**Growth and transfer:**

Epitaxial $BaTiO_3$, $SrTiO_3$ and $Sr_3Al_2O_6$ thin films were grown on $TiO_2$-terminated single-crystal (001)-oriented $SrTiO_3$ substrates via pulsed laser deposition (PLD) using a KrF excimer laser (λ = 248 nm). High-purity (99.99%, MTI Corporation) SAO targets were employed for deposition. The 10 nm thick SAO sacrificial layers were deposited at 750 °C under an oxygen pressure of $1×10^{−5}$ mbar. Subsequently, BTO (STO) films were grown on top of the SAO layer at 650 °C (700 °C) in an oxygen ambient of $5×10^{−3}$ mbar ($5×10^{−4}$ mbar). For both materials, the laser fluence was maintained at 1.8 J·cm$^{−2}$.

The growth process was monitored in situ using reflection high-energy electron diffraction (RHEED), where the intensity oscillations of the diffraction spots were used to confirm a layer-by-layer growth mode and to estimate the film thickness. To mitigate the formation of oxygen vacancies in BTO, the films underwent post-deposition annealing at 500 °C for 30 minutes in oxygen, followed by gradual cooling.

The as grown heterostructure is spin coated with PMMA (950K, 5% anisole) at 2800 rpm. The sample is scratched on the edge to remove extra PMMA and then kept in water till the sacrificial layer is dissolved. The freestanding membrane along with the polymer support is then transferred onto the silicon chip with cavities. The polymer is subsequently dissolved in acetone which results in isolated freestanding drumheads.

**PFM and KPFM measurements:**

PFM and contact KPFM measurements were performed using an NTEGRA Aura scanning probe microscope (NT-MDT, Netherlands), equipped with an external HFLI lock-in amplifier (Zurich Instruments, Switzerland) and an NI-6361 DAQ board (National Instruments Corp., USA). Platinum-coated Multi75-G probes (BudgetSensors, Bulgaria) were used in all experiments. These probes have a nominal tip radius of ~35 nm, a spring constant of approximately 5 N/m, and a typical contact resonance frequency in the range of 270–300 kHz.

All measurements were conducted inside a glovebox environment, with water and oxygen levels below 0.04 ppm. The membrane was deposited with a bottom 50-nm gold electrode, which was grounded during the experiments (see Supplementary Information Fig. S1), while the voltage was applied to the AFM probe.

PFM imaging was performed using the "chirp-type" band excitation mode(*40*), centered around the first flexural contact resonance frequency. To minimize undesired vibrations and mechanical tension in the membrane, measurements were carried out using a "jumping-mode" approach. Spectroscopy data was processed using a custom-written code in Wolfram Mathematica. Quantification of surface displacement was based on vertical and lateral force–distance curve measurements(*41*), and amplitudes were extracted using a damped harmonic oscillator model with additional normalization by the cantilever shape factor(*42*).

Contact KPFM measurements were conducted in the Dual Frequency Resonance Tracking (DFRT) mode, at the first flexural contact resonance frequency (~300 kHz). Superimposed AC and DC voltages were applied to probe the in-contact electrostatic response of the cantilever. The surface potential was determined as the DC bias corresponding to the minimum electromechanical response. The acquired data were analyzed using a custom Python script, and the lock-in amplifier phase offset was optimized to minimize the quadrature component of the hysteresis loop.

**Force response measurements:**

The force displacement measurements on the suspended membrane are performed in an AFM (Asylum Cypher) using diamond-like-carbon coated contact mode tips with a nominal spring constant of 3N/m (BudgetSensors Tap75DLC). The spring constant was further calibrated via the thermal method for the tip before measuring the *F-d* response(*43*). The cantilever displacement and its corresponding reading on the photodiode, known as the inverse optical lever sensitivity (InvOLS) was also calibrated by running a *F-d* response on the bare substrate. The origin assignment (*d* = 0) is arbitrary in all plots as the true origin of the piezo displacement has no bearing on the measurement.

**FEM Simulations:**

We performed FEM simulations with the goal of predicting the mechanical and piezoelectrical response of the membranes, and to provide a qualitative comparison to the experimental results. We employed two different types of simulations, corresponding to different sets of governing equations: nonlinear mechanical simulations of the buckling mechanism, and linear simulations of the resulting piezoelectric response. The assumption underlying such a two-step procedure is that the piezoelectric coupling has only a limited impact on the buckling process. On the contrary, the residual strain associated with the buckling drastically affects the electric response of the membrane.

**Buckling simulations**

For simulating the buckling effect, we consider the equation of linear elasticity, where a first-order constitutive relation is assumed between the stress tensor $\boldsymbol{\sigma}$ and the strain tensor $\boldsymbol{\varepsilon}$ at any point. This relation is often expressed in terms of the rank-4 stiffness tensor $\boldsymbol{C}$.

$$\boldsymbol{\sigma} = \boldsymbol{C}:\boldsymbol{\varepsilon}$$

where the "**:**" symbol denotes contraction over two indexes, i.e.:

$$\sigma_{ij} = \sum_h \sum_k C_{ijhk}\, \varepsilon_{hk}$$

Inertial forces are negligible, so the momentum conservation equation reduces to the static mechanic-equilibrium equation, i.e.:

$$\nabla \cdot \boldsymbol{\sigma} = \boldsymbol{f}$$

where $\boldsymbol{f}$ is a vector field representing the body forces, such as gravity, which in this case can also be neglected.

However, since buckling is an inherently nonlinear phenomenon, the corresponding simulations have been performed by including the geometric nonlinearity in the governing equations. This means that the momentum conservation equation is slightly modified:

$$\nabla \cdot (\boldsymbol{F}\boldsymbol{\sigma})^{\mathrm{T}} = \boldsymbol{f}$$

where the superscript "T" denotes the transposition, and $\boldsymbol{F}$ is the deformation gradient tensor. This tensor is in turn defined from the displacement field $\boldsymbol{u}$ according to the following equation

$$\boldsymbol{F} = \nabla\boldsymbol{u} + \boldsymbol{I}$$

where $\boldsymbol{I}$ denotes the identity matrix. Since we adopt additive strain decomposition, the relation between the total strain $\boldsymbol{\varepsilon}$ and its elastic and inelastic contributions, $\boldsymbol{\varepsilon}_{el}$ and $\boldsymbol{\varepsilon}_{inel}$, is simply given by:

$$\boldsymbol{\varepsilon} = \boldsymbol{\varepsilon}_{el} + \boldsymbol{\varepsilon}_{inel}$$

The abovementioned constitutive equation involving the stiffness tensor $\boldsymbol{C}$, is thus linking the elastic contribution of the strain, $\boldsymbol{\varepsilon}_{el}$, to the elastic stress $\boldsymbol{\sigma}$. Finally, the total strain is in the following relation with the displacement field:

$$\boldsymbol{\varepsilon} = \frac{1}{2}\left((\nabla\boldsymbol{u})^{\mathrm{T}} + (\nabla\boldsymbol{u}) + (\nabla\boldsymbol{u})^{\mathrm{T}}(\nabla\boldsymbol{u})\right)$$

where the third term on the right-hand side, i.e. the term quadratic in $\boldsymbol{u}$, is the one that is neglected when disregarding the geometrical nonlinearity.

We impose fixed (zero displacement) boundary conditions at the lateral borders of the membrane. Additionally, we impose an external (inelastic) in-plane strain to the whole membrane in order to simulate the residual strain that acts on the membrane and drives its buckling. The most important aspect of the model described by the equations above is that the geometrical nonlinearity causes multiple solutions to be present for given set of external conditions (i.e. boundary conditions and external strain). This is an essential feature of the buckling mechanism. It indicates that, since the phenomenon is path-dependent, the final state of a buckled membrane is not to be computed directly, but rather by employing an iterative process. In particular, we gradually increase the magnitude of the external in-plane strain while updating the solution of the static problem.

It is also important to notice that, due to the initial geometrical symmetry of the problem, the system would remain trapped in a highly unstable equilibrium state which preserves the same symmetry as that of the initial (flat) configuration. Therefore, in order to prevent this numerical artifact and force the solution to evolve along one of the stable branches, we need to superimpose a small, symmetry-breaking external load in the out-of-the-plane direction. It is also convenient that this load breaks the in-plane symmetries as well, meaning that it should not be applied along one of the symmetry-planes. This small external load has only a numerical purpose: except for the initial part of the simulation, it is much smaller than the elastic stresses that develop due to the increasing in-plane strain that is the main driver of the buckling process.

By increasing the magnitude of the in-plane strain, we can obtain more fully developed buckling patterns as solutions. In fact, doing so is equivalent to decreasing the thickness of the membrane, or to increasing the in-plane size (i.e. the size of the cavity above which the membrane is standing).

Since the thickness of the membrane is much smaller than the in-plane dimensions, the simulations have been performed using the thin shell approximation(*44*). The buckling model has been implemented and solved with the COMSOL Multiphysics software package, and specifically the Shell Interface of the Structural Mechanics Module. We used a triangular mesh composed of about $20 \times 10^3$ elements, and quadratic discretization of displacement field. We varied the magnitude of the in-plane strain by applying an auxiliary sweep to a stationary study. The magnitude has been increased with $\approx 40$ steps on a logarithmic scale. As mentioned above, we used the geometrically nonlinear formulation with additive strain decomposition. The discretized equations are then numerically integrated using the MUMPS direct solver.

The piezoelectric simulations have also been performed with COMSOL Multiphysics. Below are additional details on the corresponding governing equations and numerical implementation.

**Piezoelectric response simulations**

As mentioned, we assume that the piezoelectric response can be modelled separately, by using an input displacement field and computing the resulting electric response. The displacement field can be the output of the buckling simulations discussed above, or it may be obtained from experimental measurements of the membrane topography. In both cases, it is sufficient to consider the out-of-the-plane component of the displacement. Moreover, since the piezoelectric simulations have been performed by adopting a linear model, the magnitude of the electrical response will simply be proportional to the magnitude of the driving deformation.

When considering the piezoelectric effect, i.e. linear order electro-elastic coupling, the constitutive relation between stress and strain includes an additional coupling term when compared to the relation above:

$$\boldsymbol{\sigma} = \boldsymbol{C}:\boldsymbol{\varepsilon} - \boldsymbol{e}^\mathrm{T}\,\boldsymbol{E}$$

where $\boldsymbol{E}$ denotes the electric field, and $\boldsymbol{e}$ the rank-3 piezoelectric coupling tensor in the stress-charge form. Analogously, the constitutive relation that links the electric displacement field $\boldsymbol{D}$ with the electric field $\boldsymbol{E}$ includes a corresponding electro-elastic coupling term:

$$\boldsymbol{D} = \boldsymbol{\epsilon}\,\boldsymbol{E} + \boldsymbol{e}:\boldsymbol{\varepsilon}$$

where $\epsilon$ denotes the absolute electrical permittivity (not to be confused with the symbol "$\boldsymbol{\varepsilon}$" denoting the strain tensor). The counterpart of the static mechanical equation is the equilibrium equation of electrostatics, namely Gauss's law of electrostatics:

$$\boldsymbol{\nabla} \cdot \boldsymbol{D} = \rho_{\text{free}}$$

where $\rho_{\text{free}}$ denotes the volumetric density of free (i.e. unbound) electric charge, which in this case is zero. Similarly to the mechanical case, where the unknown variable of the problem is the displacement field $\boldsymbol{u}$, the unknown variable of the electrostatic problem is the electrical potential $V$. The analogous of the mechanical relation linking $\boldsymbol{\varepsilon}$ to $\boldsymbol{u}$ is the relation linking $\boldsymbol{E}$ to $V$, i.e.:

$$\boldsymbol{E} = -\boldsymbol{\nabla} V$$

In conclusion, the two sets of coupled equilibrium equations stated above are then solved with respect to the fields $\boldsymbol{u}(\boldsymbol{x})$ and $V(\boldsymbol{x})$. In the COMSOL environment, these governing equations correspond to the electrostatics and solid mechanics interfaces, coupled by the piezoelectricity multiphysics coupling node.

Since we are also interested in resolving stress and electric field variation through the thickness of the membrane, we did not employ the thin-shell approximation for the piezoelectric simulations. Instead, we used a tetrahedral mesh composed of $20 \times 10^3$ elements. However, it should be mentioned that in order to correctly implement the boundary conditions for the electrostatic equations, the volume surrounding the membrane needs to be included in the geometry which the electrostatics equations are applied to. The region around the membrane should be large enough for the fringing field generated by the membrane to be sufficiently small at the external boundary of the simulation region. Otherwise, the zero-charge boundary condition applied there, i.e. the normal component of $\boldsymbol{D}$ being zero, would unphysically alter the solution. This contrasts with the mechanical equations that can safely be applied only to the membrane itself. The substrate upon which the membrane is deposited (i.e. the silicon wafer) is also part of the electrostatic simulation, and it is assumed to be grounded, i.e. $V = 0$. The resolution of the mesh is not uniform throughout the geometry, but much higher in the region corresponding to the membrane itself, which is composed by about half of the total number of mesh elements of the whole geometry. The geometry used for the simulations is shown in Supplementary Information Fig. S8.

In order to obtain a smoother solution, for the piezoelectric simulations the displacement field $\boldsymbol{u}$ has been discretized using quartic element order, while quadratic discretization has been used for the electric potential $V$. For the numerical integration, the direct solver "PARDISO" has been used in this case. It should also be noted that, with the equations being linear, the use of an auxiliary sweep is not necessary for the piezoelectric simulations.

**Acknowledgement**

N.P. acknowledges funding from the ERC Advanced (NEXUS, grant no. 101054572). This work was also supported by a research grant (VIL73726) from Villum Fonden.